\documentclass[aps,prd,reprint,nofootinbib]{revtex4-2}

\usepackage{amsmath,amssymb,bm,mathtools}
\usepackage{microtype}
\usepackage{graphicx}

\usepackage{physics}   
\usepackage{tikz}
\usepackage{tikz-cd}

\usepackage[colorlinks=true,allcolors=blue]{hyperref}
\usepackage[nameinlink,capitalise]{cleveref}

\usepackage{enumitem}
\setlist[enumerate]{label=(\arabic*),leftmargin=2.2em,itemsep=0pt,topsep=2pt}

\crefname{equation}{Eq.}{Eqs.}
\Crefname{equation}{Equation}{Equations}
\crefname{figure}{Fig.}{Figs.}
\Crefname{figure}{Figure}{Figures}

\newcommand{\prlitem}[1]{\par\smallskip\noindent\textit{(#1)}\ \ignorespaces}

\begin{document}

\preprint{}

\title{Universal Nonminimal Coupling-to-Starobinsky Matching and a Single-Field Attractor}

\author{A. Sava\c{s} Arapo\u{g}lu}
\email{arapoglu@itu.edu.tr}

\author{Sermet \c{C}a\u{g}an}
\email{cagans@itu.edu.tr}

\author{Omer Guleryuz}
\email{omerguleryuz@itu.edu.tr}

\author{Cemal Berfu Senisik}
\email{senisik@itu.edu.tr}

\affiliation{Department of Physics, Istanbul Technical University, Maslak 34469 Istanbul, Türkiye}

\date{\today}

\begin{abstract}
We establish an off-shell commutativity theorem in 4D parity-even quadratic gravity that the Hubbard–Stratonovich/Legendre lifts, algebraic elimination of auxiliaries, including the torsionless Palatini connection, and Jordan–Einstein Weyl rescalings commute at the action level up to boundary terms.  This yields a frame-independent characterization of the propagating degrees of freedom and isolates a universal scalaron EFT in the metric branch, while clarifying the algebraic nature of the Palatini $f(\mathcal R)$ scalar. We obtain, as a result, a frame-universal matching from the generic nonminimal couplings to a positive $R^2$ sector and a quantitative single-field attractor bound, enhanced by a $1/\Delta N$ selection term, providing sharp and falsifiable CMB targets.
\end{abstract}

\maketitle

\section{Introduction and motivation}
Higher-derivative extensions of General Relativity (GR) provide an effective field theory (EFT) for gravity at energies below the Planck scale. The most prominent example is the Starobinsky model~\cite{Starobinsky:1980te} in which an $R+\alpha R^2$ is classically equivalent to GR plus a canonical scalar (the scalaron), as clarified by early $f(R)$ analyses based on conformal/Legendre maps~\cite{Whitt:1984pd,Maeda:1988ab,MagnanoSokolowski1994,DeFelice:2010aj}. More generally, scalar-tensor theories with algebraic $f(\phi)R$ couplings arise via Hubbard–Stratonovich (HS) lifts (auxiliary-field linearization)~\cite{Hubbard:1959ub,1957SPhD....2..416S}.

Beyond $f(R)$, Horndeski’s ghost-free interactions~\cite{Horndeski:1974wa} and Lovelock terms~\cite{Lovelock:1971yv,Clifton:2011jh} provide a broader arena to test frame changes and off-shell manipulations while avoiding Ostrogradsky instabilities~\cite{Woodard:2015zca} and respecting massive spin-2 consistency~\cite{VanNieuwenhuizen:1973fi,Hinterbichler:2011tt}. A complementary axis distinguishes the metric (Levi--Civita) and Palatini/metric--affine variational principles. For example, in the metric formulation, $R^2$ adds a scalar degree of freedom and $R_{\mu\nu}^2 / C_{\mu\nu\rho\sigma}^2$ add a massive spin-2 mode~\cite{Stelle:1976gc,Stelle:1977ry}, but in Palatini $f(\mathcal R)$ with minimally coupled matter, the extra scalar is algebraic corresponding to a Brans-Dicke parameter $\omega=-3/2$~\cite{Sotiriou:2008rp,DeFelice:2010aj,Olmo:2011uz,Flanagan:2003rb}. On the other hand, the observational CMB data strongly favor single-field Starobinsky-like inflationary dynamics with stringent limits on isocurvature perturbations and non-Gaussianity~\cite{Planck:2018jri,BICEP:2021xfz}. The volume-selection arguments further point strikingly toward single-field behavior~\cite{Tokeshi:2023swe}. This observational universality leads to the following question: \emph{Why do such diverse ultraviolet completions flow to the same infrared behavior?}

To answer such a deep question, we prove an off-shell \emph{commuting diagram} linking HS/Legendre lifts, algebraic elimination of auxiliaries (including Palatini $\Gamma$), and Weyl rescalings. Any ordering produces the same reduced action modulo the boundary terms since every step is local and algebraic~\cite{Deruelle:2010ht,Postma:2014vaa,Jarv:2014hma}. This guarantees that physical conclusions about the degrees of freedom and the effective dynamics are frame-independent. With this result, we demonstrate that the generic nonminimal couplings universally generate a positive $R^2$ sector once the heavy fields are integrated out, fixing the scalaron mass in terms of a heavy-sector projection independently of the order of the frame transformations. 

Applying the result to the early universe, we also show the combination of geometric effects associated with the Weyl scaling and a requirement that inflation persists for an observationally relevant number of e-folds. This mechanism dynamically suppresses the isocurvature perturbations, enforced by a statistical `Doob selection' mechanism, which explains the observed single-field nature of inflation~\cite{Tokeshi:2023swe} and leads to falsifiable predictions characteristic of Starobinsky-like models.

\section{Off--shell commutativity and universality}\label{SEC:II}
We consider the most general $D{=}4$ parity-even, local action quadratic in curvature invariants~\cite{Nojiri:2010wj,Myrzakulov:2014hca}:
\begin{equation}
\label{eq:quad}
S = \int d^4x \sqrt{-g} \left[
\frac{M_{\rm Pl}^2}{2}R + \alpha R^2 + \beta R_{\mu\nu}^2 + \delta C_{\mu\nu\rho\sigma}^2 + \gamma \mathcal G
\right],
\end{equation}
where $C_{\mu\nu\rho\sigma}$ is the Weyl tensor and $\mathcal G$ is the topological Gauss--Bonnet density.
Using the $D{=}4$ identities $C^2=R_{\mu\nu\rho\sigma}^2-2R_{\mu\nu}^2+\tfrac13 R^2$ and $\mathcal G=R_{\mu\nu\rho\sigma}^2-4R_{\mu\nu}^2+R^2$, the propagating sector on smooth backgrounds is controlled entirely by two linear combinations~\cite{Stelle:1976gc,Stelle:1977ry}:
\begin{equation}
\label{eq:alphap_deltap}
\alpha'=\alpha+\frac{\beta}{3},
\qquad
\delta_{\rm eff}=\delta+\frac{\beta}{2}.
\end{equation}
Physical stability requires $\alpha'>0$ (no scalar tachyon) and dictates the structure of the massive spin-2 sector governed by $\delta_{\rm eff}$. Notably, a pure $f(R)$ limit corresponds to $\beta{=}\delta{=}0$ (hence $\alpha'{=}\alpha$), while a pure Ricci-squared term ($\beta R_{\mu\nu}^2$) induces a scalar mass via the trace projection $R_{\mu\nu}^2 \supset \tfrac{1}{4}R^2$, consistent with $\alpha'=\beta/3$.

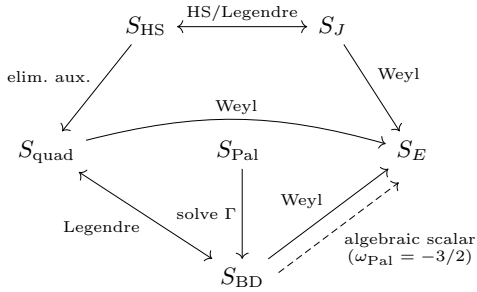
\begin{figure}[t]
\centering
\begin{tikzcd}[column sep=small]
 & S_{\rm HS} \arrow[ddl, " \text{elim.\ aux.} "'] & & S_J \arrow[ddr, " \text{Weyl} "] \arrow[ll, leftrightarrow, " \text{HS/Legendre} "']  & \\
& & & & \\
S_{\rm quad} \arrow[rrrr, bend left=15, " \text{Weyl} "] & & S_{\mathrm{Pal}}\  \arrow[dd, " \text{solve }\Gamma "'] & & S_E \\
& & & & \\
 & & S_{\rm BD} \arrow[uurr, " \text{Weyl} "]\arrow[uurr, dashed, shift right=3, "\substack{\text{algebraic scalar}\\ \text{$(\omega_{\text{Pal}}=-3/2)$}}"'] \arrow[uull, leftrightarrow, "\text{Legendre}"]& & 
\end{tikzcd}
\caption{\label{fig:commdiag} The commuting diagram of off-shell, local maps.}
\end{figure}

\prlitem{i} Commutativity theorem: We establish an off-shell \emph{commuting square} linking HS auxiliary lifts, algebraic elimination of auxiliaries (including the Palatini connection $\Gamma$), and the Jordan$\leftrightarrow$Einstein Weyl map.  Assuming auxiliary fields enter algebraically and $f''(R)\!>\!0$, these operations commute at the action level; any ordering yields the same reduced bulk action up to total--derivative boundary terms (Fig.~\ref{fig:commdiag})~\cite{Hindawi:1995an,Hindawi:1995cu,Magnano:1993bd,Flanagan:2003rb,Olmo:2011uz}.

\prlitem{ii} Spectrum and 4D spectators: The metric branch propagates a massless graviton, a scalar (scalaron) with mass $m_0^2=M_{\rm Pl}^2/(12\,\alpha')$, and a massive spin-2 ghost at $m_2^2=M_{\rm Pl}^2/(2\,\delta_{\rm eff})$.
Crucially, in $D{=}4$, $\sqrt{-g}\,C^2$ is exactly Weyl invariant, while $\mathcal G$ shifts by a total derivative. Consequently, the spin-2 sector \emph{spectates} under the Jordan$\leftrightarrow$Einstein map. This allows us to treat the ghost as a heavy regulator ($m_2 \!\gtrsim\! \Lambda_{\rm EFT}$) that decouples from the low-energy scalaron dynamics, consistent with Lee-Wick completions or causal EFT and positivity constraints~\cite{Donoghue:2019fcb,Anselmi:2018ibi,Adams:2006sv,Camanho:2014apa,Cheung:2016wjt}.

In the light-field limit ($E \ll m_2$), the system reduces to the canonical scalaron $\chi$:
\begin{equation}
S_E[\chi] = \int d^4x\sqrt{-\tilde g}\left[
\frac{M_{\rm Pl}^2}{2}\tilde R
-\frac{1}{2}(\tilde\nabla\chi)^2
- V_E(\chi)
\right],
\label{eq:EFmainboxed}
\end{equation}
where the Starobinsky potential is fixed by $\alpha'$ alone:
\begin{equation}
\label{eq:StarobinskyPot}
V_E(\chi)
=\frac{M_{\rm Pl}^4}{16\,\alpha'}\left(1-e^{-\sqrt{\frac{2}{3}}\frac{\chi}{M_{\rm Pl}}}\right)^{\!2}.
\end{equation}
Conversely, in the Palatini branch (with minimally coupled matter), the elimination of the connection $\Gamma$ yields an algebraic relation for the scalar (Brans--Dicke $\omega=-3/2$), confirming that no new propagating degrees of freedom arise~\cite{Sotiriou:2008rp,DeFelice:2010aj,Olmo:2011uz,Flanagan:2003rb}.

\section{Palatini branch: algebraic universality}\label{SEC:III}
To highlight the uniqueness of the metric scalaron, we contrast it with the Palatini (metric–affine) formulation. Here, the connection $\Gamma^\lambda_{\mu\nu}$ is treated as an independent variable, and we assume the action $S = \int \sqrt{-g} f(\mathcal{R})$ with $\mathcal{R} \equiv g^{\mu\nu}R_{\mu\nu}(\Gamma)$ and minimally coupled matter.

Introducing the auxiliary $\Phi \equiv f'(\mathcal{R})$, the equations of motion for $\Gamma$ impose the algebraic compatibility condition $\nabla^{\Gamma}_\lambda (\sqrt{-g}\,\Phi\,g^{\mu\nu})=0$. This identifies $\Gamma$ explicitly as the Levi–Civita connection of the conformal metric $\hat g_{\mu\nu} \equiv \Phi\,g_{\mu\nu}$. Consequently, the connection can be eliminated off-shell, reducing the system to a specific Brans-Dicke theory:
\begin{equation}
\label{eq:BD-omega}
S_{\rm Pal} = \int d^4x\sqrt{-g}\left[\Phi R(g) - \frac{\omega_{\rm Pal}}{\Phi}(\nabla\Phi)^2 - U(\Phi)\right],
\end{equation}
where $U(\Phi) = \Phi\mathcal{R}(\Phi)-f(\mathcal{R})$. The critical distinction is the coupling parameter:
\begin{equation}
\boxed{\quad \omega_{\rm Pal} = -\frac{3}{2} \quad}.
\end{equation}
This specific value is structurally fatal to the scalar's dynamics. Under the Weyl map to the Einstein frame, $\tilde g_{\mu\nu} = (2\Phi/M_{\rm Pl}^2)g_{\mu\nu}$, the scalar kinetic term transforms as $K_{\rm E} \propto (2\omega + 3)(\partial \ln \Phi)^2$. For $\omega_{\rm Pal} = -3/2$, this vanishes identically. The Einstein-frame action becomes purely algebraic in the scalar sector:
\begin{equation}
\label{eq:Palatini-KE-vanish}
S_E = \int d^4x \sqrt{-\tilde g} \left[ \frac{M_{\rm Pl}^2}{2}\tilde R - V_E(\Phi) \right],
\quad
V_E = \frac{M_{\rm Pl}^4 U(\Phi)}{4\Phi^2}.
\end{equation}
Thus, Palatini $f(\mathcal{R})$ propagates \emph{only} the graviton~\cite{Flanagan:2003rb,Sotiriou:2008rp}. Even with matter sources (trace $T$), the equation of motion for $\Phi$ remains a non-dynamical constraint, $2f(\mathcal{R}) - \Phi \mathcal{R} = T$, which algebraically ties $\Phi$ to the local matter density. However, this algebraicity is fragile. Relaxing the torsionless condition (allowing hypermomentum) or adding explicit $R_{\mu\nu}(\Gamma)^2$ terms generally revives the scalar or introduces ghosts by breaking the special $\omega=-3/2$ cancellation~\cite{Hehl:1994ue,Olmo:2011uz}. In this work, we focus on the metric branch where a robust, propagating scalaron naturally drives inflation.

\section{Universal $R^2$ from heavy nonminimal couplings}\label{SEC:IV}
Motivated by recent cosmological hints for nonminimal couplings~\cite{Wolf:2025jed} and their ubiquity in UV completions, we consider a generic Jordan-frame sector with $N$ scalars $\phi^I$. Decomposing the field space into a light adiabatic sector and $N{-}1$ heavy entropic directions $\chi^a$ (with mass gap $M_{\rm heavy}^2 \!\gg\! H^2$), the Lagrangian is:
\begin{equation}
\label{eq:NMCstart_prl}
\frac{\mathcal{L}_J}{\sqrt{-g}}=\frac{M_{\rm Pl}^2}{2}R + F(\phi^I)R - V(\phi^I)
-\frac{1}{2}G_{IJ}\nabla\phi^I\nabla\phi^J.
\end{equation}
We assume the heavy fields rest at a minimum $\bar\chi^a$ with a large Hessian mass matrix $(M^2)_{ab} \equiv \nabla_a \nabla_b V$.

\prlitem{i} Frame--universal matching: Fluctuations in the heavy sector $\delta\chi^a$ are sourced by the curvature via the nonminimal coupling gradient $F_a \equiv \nabla_a F$. At energies $E^2 \ll M_{\rm heavy}^2$, the kinetic terms are subleading, and the equation of motion is algebraic: $(M^2)_{ab}\delta\chi^b \simeq F_a R$. Integrating out these modes by substituting the solution back into the action generates a specific $R^2$ effective operator:
\begin{equation}
\Delta\mathcal L_{\rm eff}\supset \sqrt{-g}\ \alpha'_{\rm NMC}\,R^2,
\quad
\boxed{\,\alpha'_{\rm NMC}=\frac{1}{2}\;F_a\,(M^{-2})^{ab}F_b\,.}
\label{eq:alpha_match}
\end{equation}
Crucially, this matching is \emph{frame--universal}. Since both the heavy-field elimination (at this order) and the Weyl map are local operations, they commute. One obtains the same physical Einstein-frame scalaron mass whether one integrates out $\chi^a$ in the Jordan frame first, or Weyl-transforms first and integrates them out in the Einstein frame.

\prlitem{ii} EFT expansion and stability: The locality of the $R^2$ approximation is formally defined by the expansion of the non-local propagator $(M^2-\Box)^{-1} \simeq M^{-2} + M^{-2}\Box M^{-2}$. This generates a tower of higher-derivative operators:
\begin{equation}
\frac{\Delta\mathcal L_{\rm eff}}{\sqrt{-g}} \supset 
\alpha'_{\rm NMC} R^2
-
\underbrace{\tfrac{1}{2} F_a(M^{-2}\Box M^{-2})^{ab}F_b}_{\beta'_{\rm NMC}} R\Box R
+\dots
\end{equation}
Three physical consequences follow:
(i) \emph{Unitarity}: If the heavy sector is stable ($(M^2)_{ab}$ positive definite), then $\alpha'_{\rm NMC} \ge 0$. This guarantees that the emergent scalaron has the correct sign kinetic term (no ghost).
(ii) \emph{Covariance}: The matching result \eqref{eq:alpha_match} is built from field-space covariant tensors, ensuring invariance under redefinitions of the scalar manifold.
(iii) \emph{Observational Link}: The physical observable is the \emph{total} scalaron mass, determined by $\alpha'_{\rm tot} = \alpha'_{\rm bare} + \alpha'_{\rm NMC}$. Thus, a measurement of the tensor-to-scalar ratio $r$ (which fixes $\alpha'_{\rm tot}$) directly constrains the heavy-sector projection $F_a(M^{-2})^{ab}F_b$.

\section{Single-field attractor: geometry and selection}\label{SEC:V}
We analyze stability in the Einstein frame (EF). The Weyl map $\tilde g_{\mu\nu}=(2\Phi/M_{\rm Pl}^2)\,g_{\mu\nu}$ induces a curved field-space metric $\mathcal{G}^{(E)}_{IJ}$ and modifies the potential to $V_E = M_{\rm Pl}^4 V / (4\Phi^2)$.
Decomposing the dynamics into the adiabatic tangent $T^I$ and entropic normal $N^I$, the effective entropic mass squared $m_s^2$ governing isocurvature perturbations is
\begin{equation}
\frac{m_s^2}{H^2} = \frac{V_{E;ss}}{H^2} + 3\Omega^2 + \epsilon_H \mathcal{R}_{\rm fs},
\label{eq:ms_general_SFA}
\end{equation}
where $V_{E;ss} \equiv N^I N^J \nabla^{(E)}_I \nabla^{(E)}_J V_E$, $\Omega$ is the turn rate and $\mathcal{R}_{\rm fs}$ is the field-space curvature.
Crucially, the EF Hessian $V_{E;ss}$ receives a positive-definite contribution from the transformation itself. We identify this as the `Weyl uplift' $K^2$:
\begin{equation}
\frac{V_{E;ss}}{H^2} = 3\,\frac{V_{;ss}}{V} + K^2, \qquad K^2 \equiv \Delta_{\rm Weyl} \ge 0.
\label{eq:VEss_bridge_main}
\end{equation}
Here $V_{;ss}$ is the intrinsic Jordan-frame curvature and $K^2$ encodes the stabilizing effect of the nonminimal coupling gradients.

\prlitem{i} Selection (Doob) uplift: Beyond geometry, observational conditioning imposes a statistical
stability bound: only trajectories that remain inside the slow-roll valley until $N_F\simeq 55$ $e$-folds
contribute to the observable volume. We model the entropic displacement $s(N)$ as a 1D diffusion with linear
drift $-\mu s$ (with $\mu \equiv m_s^2/3H^2$) and diffusion coefficient $D$, with absorbing ``instability edges''
at $s=\pm\Lambda$ (where $\Lambda$ is the valley half-width). Let $h(s,N)$ denote the survival probability from
$(s,N)$ to $N_F$, i.e.\ the probability that the trajectory does not hit $|s|=\Lambda$ before $N_F$.
Conditioning is implemented by the Doob $h$-transform~\cite{Doob1957,ChetriteTouchette2015a,Risken1996},
which adds an entropic drift $2D\,\partial_s\ln h$.
For large remaining duration $\Delta N\equiv N_F-N\gg 1$ and in the bulk $|s|\ll\Lambda$, the universal long-time
\emph{bulk} dependence is Gaussian (up to an overall factor),
$h\propto \exp[-s^2/(4D\Delta N)]$~\cite{Redner2001,BrayMajumdarSchehr2013,ColletMartinezSanMartin2013},
so $\partial_s\ln h\simeq -s/(2D\Delta N)$ and hence
$2D\,\partial_s\ln h \simeq 2D\!\left(-\frac{s}{2D\Delta N}\right)= -\frac{s}{\Delta N}$ (independent of $D$).
This yields a universal additive mass shift:
\begin{equation}
\frac{m_{s,\rm sel}^2}{H^2}\simeq \frac{3}{\Delta N}.
\label{eq:Doob_mass_main}
\end{equation}
This term is not a Lagrangian parameter but a universal consequence of conditioning on sufficient inflation.

\prlitem{ii} Total attractor bound: Combining the geometric Weyl uplift with the selection pressure yields the total effective mass controlling the perturbations:
\begin{equation}
\label{eq:AttractorBound_main}
\boxed{\;
\frac{m_{s,\rm eff}^2}{H^2}
\ \equiv\ \frac{m_s^2+m_{s,\rm sel}^2}{H^2}
\ \ge\ 3\,\frac{V_{;ss}}{V} + K^2 + \frac{3}{\Delta N} \;.
}
\end{equation}

\begin{figure}[htb]
\centering
\includegraphics[width=\columnwidth]{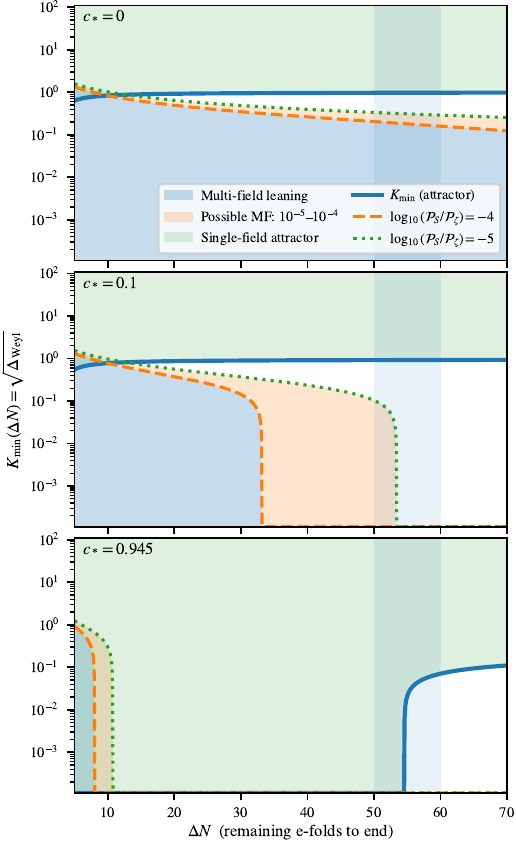}
\caption{\label{fig:sf_attractor_model} The single-field stability map. The solid blue curve marks the attractor threshold $m_{s,\rm eff}^2/H^2=1$. Regions \emph{above} this curve are safe from isocurvature due to the combined effect of the Weyl uplift ($K_{\min}$) and Doob selection ($3/\Delta N$). Dashed lines show isocurvature suppression levels ($\mathcal P_S/\mathcal P_\zeta$). Panels correspond to different baseline stabilities ($c_* \equiv 3\,V_{;ss}/V$) of the Jordan-frame potential.}
\end{figure}

If this total mass exceeds the Hubble scale ($m_{s,\rm eff} \gtrsim H$), entropic modes decouple, suppressing the power spectrum ratio $\mathcal P_S/\mathcal P_\zeta$ exponentially:
\begin{equation}
\log_{10}\frac{\mathcal P_S}{\mathcal P_\zeta}
\simeq -\frac{2}{3\ln 10}\left[
\left(\frac{m_{s,\rm eff}}{H}\right)^{\!2}\Delta N
+ 3\ln\left(\frac{\Delta N}{\Delta N_{\min}}\right)\right].
\label{eq:iso_proxy_main}
\end{equation}
For $\Delta N \approx 55$, the single-field attractor ($m_{s,\rm eff}^2/H^2 \ge 1$) requires only $3V_{;ss}/V + K^2 \gtrsim 0.945$. This creates a broad `stability window' (Fig.~\ref{fig:sf_attractor_model}) where the Weyl uplift and selection effects conspire to enforce single-field dynamics, even if the Jordan-frame potential is marginally unstable.

\section{Phenomenology: the Starobinsky target}
We focus on the metric branch predictions. Unlike the Palatini scenario (where $r \ll 10^{-3}$), the metric formulation yields a sharp target for next-generation CMB experiments, determined by the Starobinsky plateau and protected by the single-field attractor (Eq.~\eqref{eq:AttractorBound_main}).

\prlitem{i} CMB observables: Assuming the attractor ensures single-clock dynamics ($c_s\!=\!1$) and Bunch–Davies initial conditions, the observables at $N_*$ $e$-folds before the end of inflation depend solely on the plateau shape:
\begin{equation}
n_s \simeq 1-\frac{2}{N_*}, \qquad
r \simeq \frac{12}{N_*^2}, \qquad
\alpha_s \simeq -\frac{2}{N_*^2}.
\end{equation}
For the fiducial window $N_*\!\in[50,60]$, this predicts a specific tensor-to-scalar ratio $r \simeq (3.3\text{--}4.8)\times 10^{-3}$, which is squarely within the sensitivity reach of LiteBIRD~\cite{LiteBIRD:2022cnt} and CMB-S4~\cite{CMB-S4:2020lpa}.

\prlitem{ii} Probing the heavy sector: The amplitude of scalar perturbations, $\Delta_\zeta^2 \simeq 2.1\times 10^{-9}$, fixes the plateau mass scale $m_0 \simeq 10^{13}$ GeV. Since $m_0$ is determined by the \emph{total} coefficient of the $R^2$ term, a measurement of $r$ solves the `inverse problem' for the underlying theory. Using $m_0^2 = M_{\rm Pl}^2/(12\alpha'_{\rm tot})$, we find:
\begin{equation}
\alpha'_{\rm bare} + \underbrace{\frac{1}{2}F_a(M^{-2})^{ab}F_b}_{\text{Heavy Sector NMC}}
\;\simeq\; \frac{1}{24\pi^2 \Delta_\zeta^2\,r} \;\sim\; 10^9 .
\end{equation}
This is a powerful result: a detection of primordial gravitational waves ($r$) places a precise integral bound on the heavy sector's nonminimal couplings and mass spectrum, regardless of the microscopic details.

\prlitem{iii} Falsifiability: The model makes specific predictions that allow it to be ruled out. The `attractor regime' ($m_{s,\rm eff} > H$) is mandatory for the validity of the Starobinsky target. Violations of this regime would manifest as: 
\begin{enumerate}
    \item {\textit{Persistent isocurvature.}} If the attractor condition \eqref{eq:AttractorBound_main} fails, entropic modes survive, creating $\mathcal{O}(1)$ isocurvature fractions or large non-Gaussianity ($f_{\rm NL}^{\rm local} \gg 1$).
    \item {\textit{Tensor consistency violation.}} The clean prediction relies on the vacuum consistency relation $n_T = -r/8$.
    \item {\textit{Ghost modes.}} Observation of a spin-2 resonance below the EFT cutoff would invalidate the heavy-ghost assumption ($m_2 \gg H$).
\end{enumerate}
Any of these signatures would falsify the universal $R^2$ attractor described here.

\section{Discussion and outlook}
Our results find natural footing in UV completions. In string theory, curvature-squared terms descend generically from $\alpha'$ corrections, populating the 4D basis $\{R^2, R_{\mu\nu}^2, C^2\}$~\cite{Zwiebach:1985uq,Gross:1986mw,Bergshoeff:1989de,Metsaev:1987zx,Green:1984sg,Polchinski:1998rr}. Similarly, in $N{=}1$ supergravity, the $R{+}R^2$ sector is off-shell equivalent to standard SUGRA coupled to chiral matter~\cite{Cecotti:1987sa,Ferrara:2010yw}.
Crucially, the unitarity of these completions implies that the massive spin-2 ghost associated with $R_{\mu\nu}^2$ acts as a heavy regulator. By keeping $m_2 \gtrsim \Lambda_{\rm EFT} \gg H$, we ensure it decouples during inflation (satisfying Higuchi bounds~\cite{Higuchi:1986py,Deser:2001pe}), leaving behind the local operators of our effective action. The positivity of the scalaron kinetic term ($\alpha' > 0$), favored by eikonal causality~\cite{Camanho:2014apa,Adams:2006sv}, is guaranteed in our framework whenever the heavy-sector Hessian is positive definite.

Our analysis leads to three main insights. First, in the metric branch, the dynamics simplify to a compact scalaron effective field theory, whose physical mass spectrum is controlled by the shifted combinations $\alpha'=\alpha+\beta/3$ and $\delta_{\rm eff}=\delta+\beta/2$.
More broadly, we find that generic nonminimal couplings can always be matched, in a frame-independent way, onto an emergent scalaron described by $\alpha'_{\rm NMC}=\tfrac12\,F_a (M^{-2})^{ab} F_b$.  This observation has a direct observational payoff in which precision CMB measurements effectively solve an inverse problem that from the measured pair $(n_s,r)$, $\alpha'_{\rm tot}$ can be determined, which in turn constrains the projected nonminimal coupling $F_a (M^{-2})^{ab} F_b$ of the heavy-sector. When these constraints are combined with isocurvature bounds (limiting the Weyl uplift parameter $K^2$), cosmological data begin to probe the geometry of the underlying compactified manifold itself. Finally, we identify a quantitative, selection-enhanced single-field attractor bound, providing a natural explanation for the absence of observable isocurvature perturbations on CMB scales.

Thus, we proved that the reduction to the Starobinsky attractor is robust against frame changes, auxiliary elimination order, and heavy-field integration. The metric branch uniquely predicts $r \sim 10^{-3}$ and exponentially suppressed isocurvature, enforced by a statistical `Doob selection' mechanism.
A joint future confirmation~\cite{LiteBIRD:2022cnt,CMB-S4:2020lpa} of $r \approx 4\times 10^{-3}$ and the consistency relation $n_T = -r/8$ would strongly evidence this universality class. Conversely, the detection of large non-Gaussianity or persistent isocurvature would indicate a breakdown of the EFT separation or the presence of light, non-decoupled spin-2 modes.
Future work will explore the loop-induced running of $(\alpha, \beta)$ to CMB scales and the impact of hypermomentum in the Palatini branch~\cite{Olmo:2011uz,Hehl:1994ue}.

\appendix

\section{Off–Shell Commutativity Theorem}

In this Appendix, we provide the detailed derivations supporting the main text (Sec.~\ref{SEC:II} and~\ref{SEC:III}). We assume $D{=}4$ with $(-+++)$ signature and $\Phi>0$ to ensure well-defined Weyl maps. We denote total-derivative boundary contributions by $\partial (\cdots)$. Our central claim is that the linearization of higher-derivative terms (via Hubbard–Stratonovich/Legendre lifts) and Weyl rescalings commute at the action level. Table~\ref{tab:commute} summarizes the behavior of each sector under the Weyl transformation. We prove this sector by sector.

\begin{table*}[htb]
\centering
\caption{\label{tab:commute} Transformation properties of quadratic gravity sectors.}
\begin{ruledtabular}
\begin{tabular}{llll}
Sector & Transformation & Non-trivial term & Outcome \\ \hline
Scalar ($R^2$) & HS $\leftrightarrow$ Weyl & $-6\sqrt{-\tilde g}\tilde\Box\omega$ & Commutes (up to boundary) \\
Spin-2 ($C^2$) & HS $\leftrightarrow$ Weyl & None & Exact Spectator ($D{=}4$) \\
Palatini ($\Gamma$) & Elim.\ $\Gamma$ $\leftrightarrow$ Weyl & Rescaling $\hat{g} \propto \tilde{g}$ & Algebraic scalar ($\omega=-3/2$) \\
NMC & Elim.\ Heavy $\leftrightarrow$ Weyl & Pointwise extremization & Frame-universal Matching \\
\end{tabular}
\end{ruledtabular}
\end{table*}

\subsection{General Linearization Setup}

We begin with the most general parity-even action quadratic in curvature and auxiliary fields. The `parent' action containing auxiliaries $\chi$ (scalar), $\psi_{\mu\nu}$ (symmetric tensor), and $B_{\mu\nu\rho\sigma}$ (tensor) is:
\begin{align}
S = \int d^4x \sqrt{-g} \bigg[ &\tfrac{M_{\rm Pl}^2}{2}R + \gamma \mathcal{G} \nonumber \\
&+ \bar{\alpha}\left(2 \chi R - \chi^2\right) \label{eq:gen_parent} \\
&+ \bar{\beta} \left( 2 \psi_{\mu \nu} \hat{R}^{\mu \nu} - \psi_{\mu \nu}\psi^{\mu \nu} \right) \nonumber \\
&+ \bar{\delta}\left( 2 B_{\mu \nu \rho \sigma} C^{\mu \nu \rho \sigma} - B_{\mu \nu \rho \sigma}B^{\mu \nu \rho \sigma} \right) \bigg], \nonumber
\end{align}
where $\hat{R}_{\mu\nu} = R_{\mu\nu} - \frac{1}{4}Rg_{\mu\nu}$ is the traceless Ricci tensor. Varying with respect to the auxiliaries yields the algebraic equations of motion:
\begin{equation}
\chi = R, \qquad \psi_{\mu\nu} = \hat{R}_{\mu\nu}, \qquad B_{\mu\nu\rho\sigma} = C_{\mu\nu\rho\sigma}.
\end{equation}
Substituting these back into \eqref{eq:gen_parent} recovers the standard quadratic action:
\begin{equation}
S_{\rm quad} = \int d^4x \sqrt{-g} \left[ \tfrac{M_{\rm Pl}^2}{2}R + \alpha R^2 + \beta R_{\mu\nu}^2 + \delta C^2 + \gamma \mathcal{G} \right],
\end{equation}
with the identifications $\alpha = \bar{\alpha}$, $\beta = \bar{\beta}$, and $\delta = \bar{\delta}$.
Using the 4D Gauss-Bonnet and Weyl identities, the physical spectrum is determined by:
\begin{equation}
\alpha' = \alpha + \frac{\beta}{3}, \qquad \delta_{\text{eff}} = \delta + \frac{\beta}{2}.
\end{equation}

\subsection{The Scalar Channel: Proof of Commutativity}

We focus on the $f(R)$ sector (setting $\beta=\delta=0$ for clarity, implying $\alpha'=\alpha$). The action is 
\begin{equation}
    S_f = \int \sqrt{-g} f(R).
\end{equation}
\subsubsection{Route A (HS then Weyl):}
Introduce the auxiliary $\chi$ (identifying $\Phi \equiv f'(\chi)$) to write the Jordan-frame HS action:
\begin{align}
S_{\rm HS} &= \int d^4x \sqrt{-g} \Big[ \Phi R - U(\Phi) \Big], 
\\ U(\Phi) &\equiv \chi(\Phi)\Phi - f(\chi(\Phi)).
\end{align}
We now perform the Weyl rescaling $g_{\mu\nu} = e^{-2\omega}\tilde{g}_{\mu\nu}$ with $e^{2\omega} = 2\Phi/M_{\rm Pl}^2$. Using the transformation
\begin{equation}
    \sqrt{-g}\Phi R = \sqrt{-\tilde g}[\tfrac{M_{\rm Pl}^2}{2}\tilde R - \frac{3M_{\rm Pl}^2}{4\Phi^2}(\tilde\nabla\Phi)^2] + \partial (\cdots),
\end{equation}
we obtain the Einstein-frame action:
\begin{equation}
\label{eq:S1_RouteA_Result}
S_{E} = \int d^4x \sqrt{-\tilde g} \left[ \tfrac{M_{\rm Pl}^2}{2}\tilde R - \tfrac{3M_{\rm Pl}^2}{4\Phi^2}(\tilde\nabla\Phi)^2 - \tfrac{M_{\rm Pl}^4}{4\Phi^2}U(\Phi) \right].
\end{equation}

\subsubsection{Route B (Weyl then HS):}
Start with $S_f = \int \sqrt{-g} f(R)$ and perform the field-dependent Weyl rescaling $\tilde g_{\mu\nu} = \Omega^2 g_{\mu\nu}$ with $\Omega^2 = \frac{2 f'(R)}{M_{\rm Pl}^2}$. The action transforms to:
\begin{align}
S_{f} \to \int d^4x \sqrt{-\tilde g} \bigg[&\tfrac{M_{\rm Pl}^2}{2}\tilde R - \tfrac{3 M_{\rm Pl}^2}{2}(\tilde\nabla \ln f'(R))^2  \\ &- \tfrac{M_{\rm Pl}^4}{4 f'(R)^2}\big( R f'(R) - f(R) \big) \bigg].\nonumber
\end{align}
To make this local, we treat the scalar argument $R$ as an auxiliary field $\chi$. Defining $\Phi \equiv f'(\chi)$ and $U(\Phi) \equiv \chi \Phi - f(\chi)$, this becomes exactly Eq.~\eqref{eq:S1_RouteA_Result}. In conclusion,
the operations commute off-shell.

\textit{Starobinsky Potential.} For quadratic gravity, $U(\Phi) = (\Phi - M_{\rm Pl}^2/2)^2 / (4\alpha')$. Substituting this into $V_E = \frac{M_{\rm Pl}^4}{4\Phi^2}U(\Phi)$ and canonically normalizing ($\Phi = \tfrac{M_{\rm Pl}^2}{2}e^{\sqrt{2/3}\chi/M_{\rm Pl}}$) yields the potential in Eq.~\eqref{eq:StarobinskyPot} of the main text.

\subsection{The Spin-2 Channel and Spectator Status}

The spin-2 sector is governed by the traceless Ricci term. The auxiliary action is:
\begin{equation}
S_{\psi} = \int d^4x \sqrt{-g} \left[ 2\beta \psi_{\mu\nu}\left(R^{\mu\nu} - \tfrac{1}{4}Rg^{\mu\nu}\right) - \beta \psi_{\mu\nu}\psi^{\mu\nu} \right].
\end{equation}
The kinetic term for $\psi_{\mu\nu}$ has the wrong sign relative to the Einstein-Hilbert term, indicating a ghost (mass $m_2^2 = M_{\rm Pl}^2/2\delta_{\rm eff}$).

\textbf{Weyl Spectator:} In $D{=}4$, the combination $\sqrt{-g}\,C_{\mu\nu\rho\sigma}^2$ is exactly Weyl invariant. Since the massive spin-2 sector can be rewritten using the Weyl tensor (up to $R^2$ and topological terms), it \emph{spectates} under the Jordan$\to$Einstein map. The `ghost' does not mix with the scalaron dynamics during the frame change, preserving the commutative diagram.

\textit{Note on Field Redefinitions.} One can equivalently derive the spin-2 propagator by expanding the metric $g_{\mu\nu} = g^{(0)}_{\mu\nu} + h_{\mu\nu}$ and using a field redefinition $h_{\mu\nu} \to h_{\mu\nu} + \frac{4\beta}{M_{\rm Pl}^2}\psi_{\mu\nu}$ to diagonalize the quadratic action. This yields $S \sim h \mathcal{E} h - \psi (\mathcal{E} + m_2^2) \psi$, confirming the ghost nature.

\subsection{Palatini Branch: The $\omega = -3/2$ Proof}

In the Palatini formulation with torsionless $\Gamma$, the action is $S = \int \sqrt{-g} f(\mathcal{R})$ where $\mathcal{R} = g^{\mu\nu}R_{\mu\nu}(\Gamma)$.
\begin{itemize}
    \item \textbf{Legendre Lift:} $S = \int \sqrt{-g} [\Phi \mathcal{R}(\Gamma) - U(\Phi)]$.
    \item \textbf{Variation w.r.t.\ $\Gamma$:} $\nabla^\Gamma_\lambda (\sqrt{-g}\Phi g^{\mu\nu}) = 0$. This implies $\Gamma$ is the Levi-Civita connection of $\hat{g}_{\mu\nu} = \Phi g_{\mu\nu}$.
    \item \textbf{Elimination:} Using $\sqrt{-g}\Phi \mathcal{R}(\Gamma) = \sqrt{-\hat g} R(\hat g) + \partial (\cdots)$, we rewrite the action in terms of $g_{\mu\nu}$:
    \begin{equation}
    S = \int d^4x \sqrt{-g} \left[ \Phi R(g) - \frac{\omega_{\rm Pal}}{\Phi}(\partial \Phi)^2 - U(\Phi) \right],
    \end{equation}
    where $\boxed{\omega_{\rm Pal} = -\tfrac{3}{2}}$.
    \item \textbf{Einstein Frame:} Under $\tilde{g}_{\mu\nu} = \frac{2\Phi}{M_{\rm Pl}^2}g_{\mu\nu}$, the kinetic term coefficient becomes $3 + 2\omega_{\rm Pal} = 0$. The scalar disappears, leaving only the potential.
\end{itemize}

\subsection{Nonminimal Couplings (NMC)}

For the NMC sector $S = \int \sqrt{-g} [\tfrac{M_{\rm Pl}^2}{2}R + F(\phi)R - V(\phi)]$, we eliminate the fields $\phi$ algebraically. This is a Legendre-Fenchel transform:
\begin{equation}
S = \int \sqrt{-g} f_{\rm eff}(R), \;\, f_{\rm eff}(R) = \tfrac{M_{\rm Pl}^2}{2}R + \sup_{\phi} [F(\phi)R - V(\phi)].
\end{equation}
Since this operation is local, it commutes with the Weyl map. The derivative corrections (the tower of $R \Box R$ terms) arise when the heavy fields have dynamics, but the leading $R^2$ matching is frame-universal.

\section{Universal $R^2$ from Heavy Fields: Matching and Corrections}

Here we derive the effective field theory (EFT) generated by integrating out heavy scalar fields nonminimally coupled to gravity. This supports the claim in Sec.~\ref{SEC:IV} that generic UV sectors flow to the $R^2$ attractor.

\subsection{Set-up and Linearization}
Consider the generic Jordan-frame Lagrangian with $N$ scalars $\phi^I$:
\begin{equation}
\frac{\mathcal{L}_J}{\sqrt{-g}} =  \tfrac{M_{\rm Pl}^2}{2}R + F(\phi)R - \frac{1}{2}G_{IJ}(\phi)\nabla_\mu\phi^I\nabla^\mu\phi^J - V(\phi) .
\end{equation}
We decompose the fields into light (adiabatic) and heavy (entropic) directions. Let $\delta\phi^a$ ($a=1,\dots,N-1$) denote the heavy fluctuations around a background $\bar{\phi}$ where the potential is minimized. We assume a mass gap $M_{\rm heavy}^2 \gg H^2$ and slow evolution ($\dot{\bar{\phi}} \ll M_{\rm heavy}$).

Linearizing the functions $F$ and $V$ to quadratic order in $\delta\phi^a$:
\begin{align}
F(\phi) &\simeq \bar{F} + F_a \delta\phi^a, \\
V(\phi) &\simeq \bar{V} + \frac{1}{2}(M^2)_{ab} \delta\phi^a \delta\phi^b,
\end{align}
where $F_a \equiv \nabla_a F$ and $(M^2)_{ab} \equiv \nabla_a \nabla_b V$ are covariant tensors on the field space. The Lagrangian for the fluctuations is:
\begin{equation}
\frac{\mathcal{L}_{\rm heavy}}{\sqrt{-g}} \supset  -\frac{1}{2} G_{ab} \nabla \delta\phi^a \nabla \delta\phi^b - \frac{1}{2} (M^2)_{ab} \delta\phi^a \delta\phi^b + F_a \delta\phi^a R .
\end{equation}
The curvature $R$ acts as a linear source $J_a \equiv F_a R$ for the heavy fields.

\subsection{Frame-Universal Matching (Leading Order)}
At energies well below the heavy mass ($E^2 \ll M^2$), we can neglect the kinetic term $\nabla \delta\phi \nabla \delta\phi$. The equation of motion becomes algebraic:
\begin{equation}
(M^2)_{ab} \delta\phi^b \simeq F_a R \quad \Rightarrow \quad \delta\phi^a \simeq (M^{-2})^{ab} F_b R.
\end{equation}
Substituting this solution back into the action (integrating out $\delta\phi^a$ at tree level) generates the effective operator:
\begin{align}
\Delta \mathcal{L}_{\rm eff} &= \sqrt{-g} \left[ \frac{1}{2} F_a \delta\phi^a R - \frac{1}{2} \delta\phi^a (M^2)_{ab} \delta\phi^b \right] \nonumber \\ &= \sqrt{-g} \, \frac{1}{2} F_a (M^{-2})^{ab} F_b R^2.
\end{align}
Thus, we identify the generated $R^2$ coefficient:
\begin{equation}
\label{eq:S2match}
\boxed{\quad \alpha'_{\rm NMC} = \frac{1}{2} F_a (M^{-2})^{ab} F_b. \quad}
\end{equation}
\textbf{Properties:}
\begin{itemize}
    \item \textbf{Positivity:} If the heavy sector is stable ($(M^2)_{ab}$ positive definite), then $\alpha'_{\rm NMC} \ge 0$. This ensures the emergent scalaron has a healthy kinetic term.
    \item \textbf{Universality:} This matching relies only on local algebraic manipulations. It is therefore independent of whether we perform the Weyl transformation before or after integrating out the heavy fields.
\end{itemize}

\subsection{The Derivative Tower ($R \Box R$)}
To capture finite-mass effects, we retain the kinetic operator. The formal solution is $\delta\phi = (M^2 - \Box)^{-1} F R$. Expanding the propagator for large mass:
\begin{align}
(M^2 - \Box)^{-1} &= M^{-2} \left( 1 - \Box M^{-2} \right)^{-1} \nonumber \\ &\simeq M^{-2} + M^{-2} \Box M^{-2} + \mathcal{O}(M^{-6}).
\end{align}
Substituting this back yields the derivative expansion of the effective Lagrangian:
\begin{align}
\frac{\Delta \mathcal{L}_{\rm eff}}{\sqrt{-g}} =& \frac{1}{2} R \left[ F_a (M^{-2})^{ab} F_b \right] R \nonumber \\ &+ \frac{1}{2} R \left[ F_a (M^{-2} \Box M^{-2})^{ab} F_b \right] R + \dots
\end{align}
Using integration by parts ($R \Box R \to - \nabla_\mu R \nabla^\mu R$), the second term represents the leading higher-derivative correction:
\begin{equation}
\frac{\Delta \mathcal{L}_{\rm eff}}{\sqrt{-g}} \supset  \alpha'_{\rm NMC} R^2 - \beta'_{\rm NMC} R \Box R , \quad \beta'_{\rm NMC} \sim -\frac{1}{2} \frac{F^2}{M^4}.
\end{equation}
\textit{Example:} For a single heavy field $s$ with mass $m_s$ and coupling $\mu s R$, we have $\alpha' = \mu^2/(2m_s^2)$ and $\beta' = \mu^2/(2m_s^4)$.

\subsection{Einstein-Frame Geometry}
The single-field attractor analysis (Sec.~\ref{SEC:V}) relies on the Einstein-frame field-space metric. Starting from:
\begin{align}
\mathcal{L}_J &=  \sqrt{-g} \left[ \Phi(\phi) R - \frac{1}{2} G_{IJ} \nabla \phi^I \nabla \phi^J - V(\phi) \right], \\ \Phi &\equiv \tfrac{M_{\rm Pl}^2}{2} + F(\phi),
\end{align}
we apply the Weyl rescaling $\tilde{g}_{\mu\nu} = \frac{2\Phi}{M_{\rm Pl}^2}g_{\mu\nu}$.
The kinetic terms transform as:
\begin{align}
    \sqrt{-g} G_{IJ} (\nabla\phi)^2 \to& \sqrt{-\tilde g} \tfrac{M_{\rm Pl}^2}{2\Phi} G_{IJ} (\tilde\nabla\phi)^2\, \\
\sqrt{-g} \Phi R \to& \sqrt{-\tilde g} \left[ \tfrac{M_{\rm Pl}^2}{2} \tilde{R} - \tfrac{3 M_{\rm Pl}^2}{4\Phi^2} (\tilde\nabla \Phi)^2 \right] + \partial (\cdots).
\end{align}
Combining the explicit scalar kinetic term from the Weyl map ($\Phi = \Phi(\phi^I) \implies \nabla \Phi = \partial_I \Phi \nabla \phi^I$) with the original kinetic term yields the Einstein-frame metric $\mathcal{G}^{(E)}_{IJ}$:
\begin{equation}
\label{eq:S2GE}
\boxed{\quad
\mathcal{G}^{(E)}_{IJ} = \frac{M_{\rm Pl}^2}{2\Phi} G_{IJ} + \frac{3 M_{\rm Pl}^2}{2} \frac{\partial_I \Phi \partial_J \Phi}{\Phi^2}, \quad V_E = \frac{M_{\rm Pl}^4}{4\Phi^2}V.
\quad}
\end{equation}
This metric defines the geometry for the attractor bound derived in Sec.~\ref{SEC:V}.

\section{Single-Field Attractor: Geometry, Selection, and Isocurvature}
\label{secS:SFA}

We analyze the stability of the inflationary valley in the Einstein frame (EF). The effective mass of entropic fluctuations is determined by three components: the intrinsic potential curvature, the kinematic turn rate, and the geometric `Weyl uplift' induced by the frame transformation.

\subsection{Einstein-Frame Kinematics}
Let $T^I \equiv \dot\phi^I/\dot\sigma$ be the adiabatic unit tangent and $N^I$ the entropic unit normal with respect to the Einstein-frame metric $\mathcal{G}^{(E)}_{IJ}$. The background speed is $\dot\sigma = \sqrt{2\epsilon_H} H M_{\rm Pl}$. The dimensionless turn rate is $\Omega \equiv \|D_t T\|/H$.
At leading order in slow-roll, the effective mass squared for the entropic mode $s$ is:
\begin{equation}
\label{eq:S3:msmaster}
\frac{m_s^2}{H^2} \;=\; \frac{V_{E;ss}}{H^2} \;+\; 3\Omega^2 \;+\; \epsilon_H \mathcal{R}_{\rm fs},
\end{equation}
where $V_{E;ss} \equiv N^I N^J \nabla^{(E)}_I \nabla^{(E)}_J V_E$ and $\mathcal{R}_{\rm fs}$ is the field-space Ricci scalar.
Since $3\Omega^2 \ge 0$ and $\epsilon_H \ll 1$, a sufficient condition for stability is $V_{E;ss}/H^2 \gtrsim 1$.

\subsection{The Weyl Uplift ($K^2$)}
The Einstein-frame potential $V_E = M_{\rm Pl}^4 V / (4\Phi^2)$ mixes the original curvature with the gradients of the conformal factor. Calculating the covariant Hessian and projecting onto the normal direction yields the decomposition:
\begin{equation}
\label{eq:VEss_bridge_main}
\boxed{\quad
\frac{V_{E;ss}}{H^2} = 3\,\frac{V_{;ss}}{V} + K^2.
\quad}
\end{equation}
Here, $V_{;ss}$ is the purely Jordan-frame curvature term. The second term, which we call the \textbf{Weyl Uplift}, is explicitly positive-definite:
\begin{align}
\label{eq:S3:DeltaW_def}
K^2 &\equiv \Delta_{\rm Weyl}\nonumber  \\ &= 3 \left[ 4(\partial_s \ln\Phi)^2 - 2\nabla_s \nabla_s \ln\Phi - 4(\partial_s \ln\Phi)\frac{V_{;s}}{V} \right].
\end{align}

This geometric correction $K^2 \ge 0$ implies that the Einstein-frame valley is generically \emph{steeper} than its Jordan-frame counterpart, enhancing stability for nonminimally coupled theories.

\subsection{Doob Selection and the Effective Mass}

Inflationary trajectories are conditioned on survival: they must not exit the slow-roll valley before
$N_F \approx 60$ $e$-folds. We model the transverse fluctuations $s(N)$ as an Ornstein--Uhlenbeck process
(with approximately constant coefficients during slow roll) with absorbing boundaries at $\pm\Lambda$,
\begin{equation}
\frac{ds}{dN} = -\mu\, s + \sqrt{2D}\,\eta(N),
\qquad
\mu \equiv \frac{m_s^2}{3H^2},
\label{eq:S3_OU_SDE}
\end{equation}
where $\langle \eta(N)\eta(N')\rangle=\delta(N-N')$ and $D$ is the diffusion coefficient.
Let $h(s,N)\equiv \mathbb{P}(\tau_\Lambda>N_F\,|\,s(N)=s)$ denote the survival probability up to $N_F$,
with $\tau_\Lambda$ the first hitting time of $\pm\Lambda$. Then $h$ obeys the backward Fokker-Planck equation
\cite{Risken1996}
\begin{equation}
-\partial_N h
=
\Big(D\,\partial_s^2 - \mu s\,\partial_s\Big) h,
\qquad
h(\pm\Lambda,N)=0,
\label{eq:S3_backward_FP}
\end{equation}
where $h(s,N_F)=1 \ (|s|<\Lambda)$. Then, conditioning on survival is implemented by the Doob $h$-transform~\cite{Doob1957,ChetriteTouchette2015a},
which modifies the generator as $L_b^{(h)}f = h^{-1}L_b(hf)$ and, equivalently, adds an entropic drift.
At the level of the Langevin equation, the conditioned process reads
\begin{equation}
\frac{ds}{dN}
=
-\mu s
+ 2D\,\partial_s \ln h(s,N)
+ \sqrt{2D}\,\eta(N).
\label{eq:S3_Doob_drift}
\end{equation}
This makes explicit that conditioning changes the dynamics at the level of the stochastic generator,
independently of any Lagrangian parameters.

To extract the universal late-time behavior, define the remaining duration $\Delta N \equiv N_F-N$ and work
in the bulk regime $|s|\ll \Lambda$ with $\Delta N\gg 1$. In this regime the universal selection effect is
captured by momentarily dropping the weak OU drift in the backward equation, i.e.\ setting $\mu\simeq 0$ in
Eq.~\eqref{eq:S3_backward_FP}, which reduces it to the heat equation
\begin{equation}
\partial_{\Delta N} h \simeq D\,\partial_s^2 h .
\label{eq:S3_heat_equation}
\end{equation}
Its large-$\Delta N$ bulk solution (the universal ``meander'' profile) is Gaussian~\cite{Redner2001}:
\begin{equation}
h(s,\Delta N)\propto \exp\!\left[-\frac{s^2}{4D\Delta N}\right]
\quad \Rightarrow \quad
\partial_s \ln h \simeq -\frac{s}{2D\Delta N}.
\label{eq:S3_meander_profile}
\end{equation}
Therefore the Doob term becomes
\begin{equation}
2D\,\partial_s \ln h \simeq -\frac{s}{\Delta N}.
\label{eq:S3_Doob_force}
\end{equation}
Crucially, the diffusion coefficient cancels: the selection-induced restoring force is universal.
Thus the effective linear drift is $-(\mu + 1/\Delta N)s$, which is equivalent to an additive mass shift.
Defining $\mu_{\rm sel}\equiv 1/\Delta N$, we obtain the \textbf{Selection Mass}
\begin{equation}
\boxed{\quad
\frac{m_{s,\rm sel}^2}{H^2}
\equiv 3\mu_{\rm sel}
\simeq \frac{3}{\Delta N}\,.
\quad}
\label{eq:S3_selection_mass}
\end{equation}

Combining the geometric Weyl uplift with the selection pressure yields the total effective mass bound:
\begin{equation}
\frac{m_{s,\rm eff}^2}{H^2}
\equiv
\frac{m_s^2+m_{s,\rm sel}^2}{H^2}
\ \ge\
3\frac{V_{;ss}}{V} + K^2 + \frac{3}{\Delta N}.
\label{eq:S3_total_bound}
\end{equation}

\subsection{Isocurvature Suppression}
On super-horizon scales, the isocurvature mode $S$ decays as $dS/dN = -(m_{s,\rm eff}^2/3H^2)S$. Integrating this from horizon crossing to the end of inflation yields the power suppression:
\begin{equation}
\log_{10} \frac{\mathcal{P}_S}{\mathcal{P}_\zeta} \simeq -\frac{2}{3\ln 10} \left[ \left(\frac{m_s}{H}\right)^2 \Delta N + 3 \ln \left(\frac{\Delta N}{\Delta N_{\rm min}}\right) \right].
\end{equation}
This formula generates the dashed `isocurvature ceiling' contours in Fig.~\ref{fig:sf_attractor_model}.

\subsection{Example: Linear NMC on a Starobinsky Plateau}
\begin{figure}[htb]
\centering
\includegraphics[width=0.95\columnwidth]{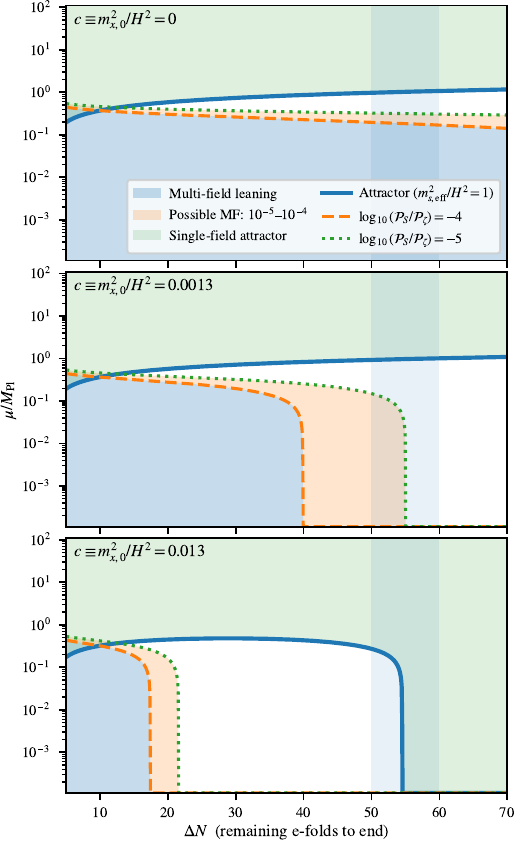}
\caption{\label{fig:sf_attractor_model_starobinsky} Stability map for the Linear-NMC Starobinsky model. The vertical axis is the NMC strength $\mu$, and the horizontal axis is remaining $e$-folds $\Delta N$. Regions above the solid blue line ($m_{s,\rm eff}^2/H^2 = 1$) are dynamically stable single-field attractors. The dotted contours show isocurvature suppression levels.}
\end{figure}
To illustrate these bounds, consider a specific model with adiabatic field $\varphi$ and a heavy field $x$ with linear nonminimal coupling:
\begin{equation}
\Phi = \tfrac{M_{\rm Pl}^2}{2} + F(\varphi) + \mu x, \quad V_J = \frac{F(\varphi)^2}{4\alpha'} + \left(\frac{2\Phi}{M_{\rm Pl}^2}\right)^2 \frac{1}{2}m_{x,0}^2 x^2.
\end{equation}
In the Einstein frame, this factorizes into a Starobinsky potential for $\chi$ and a quadratic valley for $x$.
The intrinsic Hessian ratio is $c \equiv m_{x,0}^2/H^2$. Evaluating the terms at the valley bottom ($x=0$):
\begin{equation}
3\frac{V_{;ss}}{V} \simeq \frac{4}{3}c \Delta N, \qquad K^2 \simeq \frac{54\mu^2}{M_{\rm Pl}^4 \Delta N}.
\end{equation}
(Using the plateau approximation $\Phi \propto \Delta N$).
This leads to a sufficient condition for stability ($m_{s,\rm eff}^2/H^2 \ge 1$) at $\Delta N = 55$:
\begin{equation}
\label{eq:LinearNMC_Bound}
c \gtrsim \frac{3}{220} \left( 0.945 - \frac{36\mu^2}{M_{\rm Pl}^2} \right).
\end{equation}
Figure~\ref{fig:sf_attractor_model_starobinsky} plots this stability map. A non-zero coupling $\mu$ enhances stability via the Weyl uplift $K^2$, allowing for lighter intrinsic masses $c$ while preserving single-field dynamics.

\bibliography{apssamp}

\end{document}